\documentclass{emulateapj}

\slugcomment{9/27/12 for AJ}
\shorttitle{Radio Nucleus in NGC\,3115}
\shortauthors{Wrobel \& Nyland}
\begin{document}
\title{Discovery of a Flat-Spectrum Radio Nucleus in NGC\,3115}
\author{J. M. Wrobel\altaffilmark{1,2} and K. Nyland\altaffilmark{3}} 

\altaffiltext{1}{National Radio Astronomy Observatory, P.O. Box O,
  Socorro, NM 87801; jwrobel@nrao.edu}

\altaffiltext{2}{The National Radio Astronomy Observatory (NRAO) is a
  facility of the National Science Foundation, operated under
  cooperative agreement by Associated Universities, Inc.}

\altaffiltext{3}{New Mexico Tech, Department of Physics, 801 Leroy
  Place, Socorro, NM 87801;knyland@nrao.edu}

\begin{abstract}
The early-type galaxy NGC\,3115, at a distance of 10.2 Mpc, hosts the
nearest billion-solar-mass black hole. Wong et al. recently inferred a
substantial Bondi accretion rate near the black hole. Bondi-like
accretion is thought to fuel outflows, which can be traced through
their radio emission. This paper reports the discovery of a radio
nucleus in NGC\,3115, with a diameter less that 0.17\arcsec\,
(8.4~pc), a luminosity at 8.5~GHz of $3.1 \times
10^{35}$~ergs~s$^{-1}$ and a flat spectrum ($\alpha = -0.23\pm0.20$,
$S \propto \nu^\alpha$). The radio source coincides with the galaxy's
photocenter and candidate X-ray nucleus. The emission is radio-loud,
suggesting the presence of an outflow on scales less than 10~pc. On
such scales, the Bondi accretion could be impeded by heating due to
disruption of the outflow.
\end{abstract}


\section{Motivation}

Studies of the nearest massive black holes (MBHs) offer unique
insights into inflows and outflows in galactic nuclei. Such MBHs can
be investigated on a key, physically-defining scale. Spherically
symmetric, adiabiatic accretion is characterized by a Bondi radius
$R_{\rm B} = 2 G M_\bullet / c^2_s$ and accretion rate $\dot{M}_{\rm
  B} = \pi R^2_B \rho c_s$, where $M_\bullet$ is the MBH mass, and
$\rho$ and $c_s$ are the density and sound speed of the thermal gas at
$R_{\rm B}$. For the nearest galaxies $R_{\rm B} \sim 100$~pc, which
can be spatially resolved by {\em Chandra\/} and allows $\dot{M}_{\rm
  B}$ to be inferred \citep[e.g.,][]{pel05}.

Jet-like outflows are a hallmark of low-luminosity active galactic
nuclei (AGNs) \citep[][and references therein]{nag05,ho08} and are
thought to be fueled by Bondi-like inflows \citep[e.g.,][]{pel05}. On
scales of order $R_{\rm B}$ or more, such outflows can be traced by
radio imaging of jets and lobes, and X-ray imaging of jet-induced
cavities in the thermal gas \citep[e.g.,][]{all06,mcn07}. On scales
much less than $R_{\rm B}$, such outflows have only been traced using
radio imaging \citep[][and references therein]{nag05}.

This work focuses on the nearest billion-solar-mass BH, that hosted by
the flattened early-type galaxy NGC\,3115 \citep{kor96,ems99}. A
distance of $D = 10.2$ Mpc, a nominal BH mass of $M_\bullet =
9.6\times10^8 M_\odot$ and a 1 $\sigma$ mass range of $M_\bullet = 6.7
- 15 \times10^8 M_\odot$ \citep{ems99,gul09a} are adopted. For a
canonical radiative efficiency of $\eta = 0.1$, the associated
Eddington luminosity is a quasar-like $L_{\rm Edd} = 1.2 \times
10^{47} $~ergs~s$^{-1}$. The MBH is embedded in a nuclear star cluster
located at the photocenter of NGC\,3115's stellar disk and cuspy bulge
\citep{kor96,ems99,lau05}.  Table~1 gives the position of the galaxy
photocenter.

NGC\,3115 is hot-gas poor and its overall gas temperature, 0.4 keV, is
atypically high compared to early-type galaxies with similar hot-gas
contents \citep{bor11}. \citet{won11} report that the gas temperature
is 0.3 keV in a 4\arcsec - 10\arcsec\, annulus and begins to rise at a
radius of 4\arcsec - 5\arcsec. This was interpreted as evidence for
spatially resolving the Bondi flow onto the MBH, with the flow
characterized by $R_{\rm B} = 4\arcsec - 5\arcsec$ (200-250~pc) and
$\dot{M}_{\rm B} = 2.2 \times 10^{-2}$ M$_\odot$ yr$^{-1}$. This Bondi
radius is consistent with expectations given the uncertainty in the BH
mass \citep{gul09a}.

A basic tenet of Bondi accretion theory is that the flow is adiabatic,
a situation that could be violated if heating mechanisms are in play
near the MBH. Here, we set the stage for examining the role of
outflows that could potentially heat the flow toward the MBH in
NGC\,3115. We present new radio constraints on any outflows
(\S~\ref{imaging}) and interpret those contraints within the context
of other information on NGC\,3115 (\S~\ref{implications}). We close in
\S~\ref{summary} with a summary and conclusions.

\section{Imaging}\label{imaging}

NGC\,3115 was observed with the Very Large Array \citep[VLA;][]{tho80}
in its A configuration on 2004 November 16 UT under proposal code
AT299. A coordinate equinox of 2000 was employed. J1007-027, at a
position of $\alpha(J2000) = 10^{h} 07^{m} 04\fs3499$, $\delta(J2000)
= -02\arcdeg 07\arcmin 10\farcs917$ and with a one-dimensional
position error at 1 $\sigma$ better than 1 mas, was used as a phase
calibrator. The switching time between it and NGC\,3115 was 490~s,
with a switching angle of 5.6\arcdeg. To avoid phase-center artifacts,
the {\em a priori\/} pointing position for NGC\,3115 was about
0.2\arcsec\, South of the then-available {\em Chandra\/}
position. Data were acquired in dual circular polarizations with a
bandwidth of 0.1~GHz centerd at 8.4601~GHz (8.5 GHz
hereafter). Observations of 3C\,286 were used to set the amplitude
scale to an accuracy of about 3\%. The net exposure time on NGC\,3115
was 5180~s.

The data were calibrated and imaged using release 3.3.0 of the Common
Astronomy Software Applications (CASA) package
\citep{mcm07}. Twenty-five of 27 antennas provided data of acceptable
quality, with most of the data loss due to EVLA \citep{per11}
retrofitting activities. The CASA task {\tt clean} was used to form
and deconvolve a naturally-weighted image of the Stokes $I\/$ emission
from NGC\,3115. NRAO's Astronomical Image Processing System (AIPS)
\citep{gre03} was used for image analysis. Figure~1 shows the image,
which has an rms noise level of $\sigma = 0.018$~mJy~beam$^{-1}$. A
search within the NED/2MASS error circle (Table~1) led to the
detection of one source. A Gaussian fit to that source in the image
plane was made using the AIPS task {\tt JMFIT}. From this fit, the
source was found to be compact with a diameter less than 0.17\arcsec\,
and with an integrated flux density of $S_{\rm 8.5~GHz} =
0.29\pm0.03$~mJy, where the quoted error is the quadratic sum of the
3\% scale error and the error in the fit. The fit also yielded a
position at 8.5 GHz with an error dominated by that due to the
phase-referencing strategies (Table~1, Fig.~1). With adequate
signal-to-noise, structures as large as 6\arcsec\, could be
represented in Figure~1. No extended structure was found.

A VLA image of NGC\,3115 at 1.4 GHz is also available from the Faint
Images of the Radio Sky at Twenty centimeters (FIRST) survey of
\citet{whi97}. Near the 8.5 GHz position there is a weak detection at
1.4 GHz obtained on 2002 August 12 UT with a geometric resolution of
5.9\arcsec. Because of the source's weakness, a parabolic fit in the
image plane was made using the AIPS verb {\tt MAXFIT}. This fit
yielded a peak flux density of $S_{\rm 1.4~GHz} =
0.44\pm0.15$~mJy~beam$^{-1}$, only 2.9 times the local rms noise
level. The fit also provided a position at 1.4 GHz with an error
dominated by that due to the signal-to-noise ratio (Table~1, Fig.~1).
NGC\,3115 was also observed briefly at 1.4 GHz under proposal code
AT299; those data were calibrated and imaged but that image is not
presented here because it had poorer sensitivity than the FIRST image.

\section{Implications}\label{implications}

\subsection{The Radio Nucleus}

From Figure~1, the 8.5 GHz detection of NGC\,3115 has a diameter less
than 0.17\arcsec\, (8.4~pc). Its flux density is consistent with the 3
$\sigma$ upper limit of 0.33~mJy measured at 4.9 GHz by \citet{fab89}
with a geometric resolution of 8.5\arcsec\, (420~pc), and thus on the
Bondi-zone scales \citep{won11} shown in Figure~1. Such photometric
consistency implies that Figure~1 is recovering almost all of the
high-frequency emission within the Bondi zone. The photometry being
compared at 4.9 and 8.5 GHz is separated by 17 years. If this
comparison is compromised by time variability, the case for the
compact nature of the high-frequency emission would be
strengthened. The available high-frequency data provide no evidence
for extended radio emission within the Bondi zone. In this
circumstance the spectral index between 8.5 and 1.4 GHz can be
usefully computed: it is $\alpha = -0.23\pm0.20$ ($S \propto
\nu^\alpha$). This index could be compromised by time variability over
about two years but, again, such variability would strengthen the case
for the compact nature of the radio emission.

The compact, flat-spectrum radio source coincides astrometrically with
NGC\,3115's photocenter (Table~1, Fig.~1). Because of this astrometric
agreement, Table~1 refers to this source as the radio nucleus of
NGC\,3115. Its luminosity is $\nu L_{\nu}(8.5~GHz) = 3.1 \times
10^{35}$~ergs~s$^{-1}$.

\subsection{The Candidate X-ray Nucleus}

\citet{zha09} identified a pointlike {\em Chandra\/} source close to
the galaxy's photocenter (and thus its MBH) and surrounded by diffuse
X-ray emission. Other studies also suggest a pointlike X-ray source
\citep{ho09,gul09b,bor11,mil12}. After correction for stellar and
thermal emission, \citet{bor11} find that the pointlike source has a
hard spectrum, a 2-10 keV luminosity of $L_{\rm X} = 4.3 \times
10^{38}$~ergs~s$^{-1}$ and an Eddington fraction of $L_{\rm X} /
L_{\rm Edd} = 3.6 \times 10^{-9}$. Table~1 gives the position of this
nuclear X-ray source, CXO J100513.9-074307, from Release 1.1 of the
{\em Chandra\/} Source Catalog \citep{eva10}. However, both
\citet{won11} and \citet{mil12} note possible blending issues that
could affect both the astrometry and photometry of this source. For
this reason, the cited X-ray luminosity and Eddington fraction are
conservatively treated as upper limits ($L_{\rm X} < 4.3 \times
10^{38}$~ergs~s$^{-1}$, $L_{\rm X} / L_{\rm Edd} < 3.6 \times
10^{-9}$) and Table 1 refers to this source as the candidate X-ray
nucleus of NGC\,3115.

\subsection{Potentially Heating the Inflow}

The compact, flat-spectrum radio source coincides astrometrically with
NGC\,3115's candidate X-ray nucleus (Table~1, Fig.~1). A study of
radio and X-ray sources in low-luminosity AGNs found that the majority
are radio loud, defined as $log~R_X = log~\nu L_{\nu}(5~GHz) / L_{\rm
  X} = -4.5$ or higher \citep{ter03}. That study involved radio
sources with flat or inverted spectra, so the cited definition applies
equally well at 8.5~GHz as at 5~GHz. For NGC\,3115, $log~R_X = log~\nu
L_{\nu}(8.5~GHz) / L_{\rm X} > -3$, implying it is radio-loud. As
conservatively estimated above, the Eddington fraction of the MBH in
NGC\,3115 is $L_{\rm X} / L_{\rm Edd} < 3.6 \times 10^{-9}$. This is
remarkably small given that \citet{won11} find that the Bondi inflow
onto the MBH is characterized by a substantial accretion rate of
$\dot{M}_{\rm B} = 2.2 \times 10^{-2}$ M$_\odot$ yr$^{-1}$ at a radius
of $R_{\rm B} = 4\arcsec - 5\arcsec$ (200-250~pc). Could this inflow
be somehow impeded, limiting the fuel supplied to the MBH? Indeed, the
density slope inferred by \citet{won11} led them to suggest that the
accretion was being suppressed near the MBH.

The compact, flat-spectrum and radio-loud nature of many
low-luminosity AGNs has been demonstrated to arise from outflows on
parsec scales \citep[][and references therein]{nag05,ho08}. The radio
nucleus of NGC\,3115 is compact, has a flat spectrum and is
radio-loud, so it could plausably support an outflow on parsec
scales. Mechanical feedback from such an outflow could heat, and thus
impede, the Bondi inflow characterized by \citet{won11}. To be most
effective, the parsec-scale outflow should be poorly collimated,
curved and/or bent, traits often seen among low-luminosity AGNs
\citep{nag05}. For example, the iconic low-luminosity AGN in the
early-type galaxy NGC\,4278 has long been known to feature
flat-spectrum and radio-loud emission on parsec scales \citep[][and
  references therein]{gir05}. Notably, \citet{pel12} recently
concluded that the distorted, parsec-scale jets in NGC\,4278 play a
key role in heating the ambient medium in this early-type galaxy. Our
discovery of compact, flat-spectrum and radio-loud emission from
NGC\,3115 sets the stage for future radio imaging using very long
baseline interferometry (VLBI). Knowing the radio structure of
NGC\,3115 on parsec scales, its potential for heating the ambient
medium can be quantified.

It is also possible that the inflow is heated by stellar motions in
NGC\,3115's parsec-scale nuclear cluster, for which the innermost
velocity dispersion reaches about 600 km~s$^{-1}$
\citep{kor96}. \citet{hil12} recently presented a theoretical study of
this effect. Such stellar-based heating could also influence the
formation and propagation of outflows driven by the MBH in NGC\,3115.
For example, if it is difficult for the outflow to propagate then VLBI
imaging could show emission confined to sub-parsec scales.

\subsection{Radio Emission from an Outflow or Inflow?}

A word of caution is in order. The flat-spectrum radio source in
NGC\,3115 has an 8.5 GHz luminosity that is only about 290 times that
of the flat-spectrum source Sagittarius A$^\star$, the radio nucleus
of the Milky Way \citep[][and references therein]{gen10}. That nucleus
is about eight orders of magnitude underluminous compared to its
Eddington luminosity. There is an ongoing debate about whether the
steady radio emission from Sagittarius A$^\star$ traces an outflow,
arises from the accretion flow itself, or involves both phenomena
\citep[][and references therein]{yua11}. Thus, at these low radio
luminosities, outflows could be entirely absent and unavailable for
heating. Structural information on parsec scales could help
distinquish between these two scenarios for NGC\,3115, further
underscoring the need for VLBI imaging. Such imaging could reveal
either elongated structures that are outflow driven or pointlike
emission from the accretion flow near the MBH. In the interim, the
radio photometry presented here can help constrain models of the
emission from the outflows and/or inflows in the vicinity of the
billion-solar-mass BH in NGC\,3115.

\section{Summary and Conclusions}\label{summary}

We analyzed archival 8.5~GHz VLA observations from the A-configuration
of the nucleus of NGC\,3115 and detected, for the first time, compact
emission (diameter $<$ 8.4~pc) with a luminosity of $3.1 \times
10^{35}$~ergs~s$^{-1}$. The compact radio emission in NGC\,3115 is
spatially coincident with the optical center of the galaxy as well as
the candidate X-ray nucleus. In addition to the 8.5~GHz detection, we
found a weak detection in the FIRST image at 1.4~GHz and showed that
the emission has a flat radio spectrum, with a spectral index of
$\alpha = -0.23 \pm 0.20$.  We compared the 8.5~GHz and 2-10 keV
luminosities and found that NGC\,3115 is radio-loud, with $log~R_{X} >
-3$.

\citet{won11} recently reported a {\em Chandra\/} detection of a
centrally-rising gas temperature gradient in the nucleus of NGC\,3115,
and cited this detection as evidence of spatially resolved Bondi
accretion onto the MBH on scales of 200-250~pc with $\dot{M}_{\rm B} =
2.2 \times 10^{-2}$ M$_\odot$ yr$^{-1}$. However, this substantial
Bondi accretion rate is at odds with the extremely low Eddington
fraction of $L_{\rm X} / L_{\rm Edd} < 3.6 \times 10^{-9}$, suggesting
that heating within the nucleus of NGC\,3115 may be impeding the Bondi
inflow. As has been established in other low-luminosity AGNs, the
radio-loud nature of NGC\,3115 could be the result of a parsec-scale
outflow capable of heating the ambient gas and starving the central
MBH. The high-velocity stellar motions in the parsec-scale nuclear
star cluster in NGC\,3115 also have the potential to heat the central
gas, possibly hindering the accretion of material onto the MBH and
confining any outflows to sub-parsec scales. Alternatively, pointlike
radio emission could also arise from the accretion flow itself.
Future, sensitive VLBI observations are needed to distinguish among
these possibilities.

\acknowledgements 
This research has made use of data obtained from the {\em Chandra\/}
Source Catalog, provided by the {\em Chandra\/} X-ray Center (CXC) as
part of the {\em Chandra\/} Data Archive. We are grateful to Yuichi
Terashima, Luis Ho and Jim Ulvestad for their early efforts in
proposing and observing AT299.

{\it Facilities:} \facility{{\em Chandra}}, \facility{VLA}.

\clearpage

\begin{figure}
\plotone{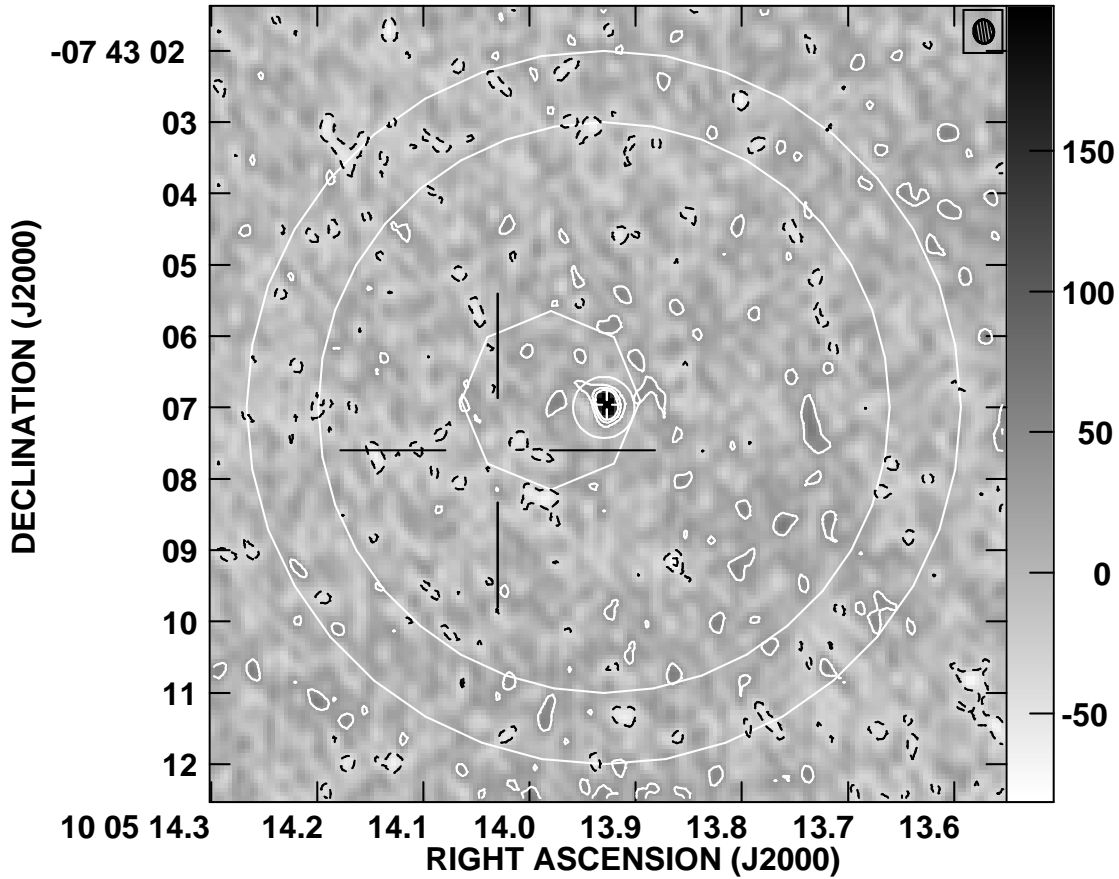}
\caption{VLA image of Stokes $I\/$ emission from NGC\,3115 at a
  frequency of 8.5 GHz and spanning 11\arcsec\, (540 pc). The symbols
  mark positions and their errors at the 95\% confidence level
  (Table~1). Symbols are an octagon for the NED/2MASS position, a
  small circle for the candidate X-ray nucleus, a large cross with a
  gap for the 1.4 GHz nucleus; and a small cross with a gap for the
  8.5 GHz nucleus. In addition, the large circles encode the range in
  the Bondi radius from Wong et al. (2011). For the 8.5 GHz image,
  natural weighting was used, giving an rms noise of
  0.018~mJy~beam$^{-1}$ (1 $\sigma$) and beam dimensions at FWHM of
  0.33\arcsec\, $\times$ 0.27\arcsec\, with elongation PA = 9\arcdeg\,
  (hatched ellipse). Allowed contours are at -6, -4, -2, 2, 4, and 6
  times 1 $\sigma$. Negative contours are dashed and positive ones are
  solid.  Linear grey scale spans -0.08~mJy~beam$^{-1}$ to
  0.20~mJy~beam$^{-1}$. Scale is 1\arcsec\, = 49.5 pc}\label{fig1}
\end{figure}

\begin{deluxetable}{lcccc}
\tabletypesize{\scriptsize}
\tablecolumns{5}
\tablewidth{0pc}
\tablecaption{Astrometry of Nuclear Components}\label{tab1}
\tablehead{
\colhead{}          & \colhead{R.A.}    & \colhead{Decl.}   & 
\colhead{Error}     & \colhead{}\\
\colhead{Component} & \colhead{(J2000)} & \colhead{(J2000)} &
\colhead{(\arcsec)} & \colhead{Ref.}\\
\colhead{(1)}       & \colhead{(2)}     & \colhead{(3)}     &
\colhead{(4)}       & \colhead{(5)}}
\startdata
2 $\mu$m galaxy photocenter & 10 05 13.98  & -07 43 06.9  & 1.25 & 1\\
Radio nucleus, 8.5 GHz      & 10 05 13.927 & -07 43 06.96 & 0.20 & 2\\
Radio nucleus, 1.4 GHz      & 10 05 14.03  & -07 43 07.6  & 2.2  & 2\\
Candidate X-ray nucleus     & 10 05 13.93  & -07 43 07.0  & 0.43 & 3\\
\enddata
\tablecomments{
Col.~(1): Component.
Cols.~(2) and (3): Component position. Units of right ascension are 
hours, minutes, and seconds, and units of declination are degrees, 
arcminutes, and arcseconds.
Col.~(4): Diameter of error circle at 95\% confidence level.
Col.~(5): Reference.}
\tablerefs{(1) NED/2MASS; (2) this work; (3) Evans et al. 2010,
  Release 1.1.}
\end{deluxetable}

\end{document}